\documentclass[12pt]{article}
\usepackage{graphicx}

\begin{document}

\title{Large Radio Astronomy: next 70 Years Step}

\author{Parijskij Yu.N.\thanks{par@sao.ru}}
\date{}
\maketitle
\begin{center}
Special Astrophysical Observatory of the Academy of Sciences of
Russia, Nizhnij Arkhyz, Russia
\end{center}

\begin{abstract}
Some attempts to predict the very distant future of Radio
Astronomy are given. It is not easy to predict a list of the first
priority problems which may appear, but the facilities potential
is more predictable. It is suggested, that in addition to the
"dedicated for Radio Astronomy", facilities may be extended
greatly by integration with the next generation living standards
facilities, connected with People-to-People communications through
the global networks and by incorporating of  the "Natural
facilities", such as grav. lensing,  maser amplification in the
ISM etc. As an examples of the extreme cases of the  $10^9 m^2$
class of the new generation Radio Telescopes, utilization of the
personal dipole size communication facilities by SKA type
instrument, and array from the asteroids first "Frehnel zones"
will be mentioned. Radio Astronomy from the secondary to optical
facilities tool will be the only tool in the exploration of the
$z>10$ Universe. The reality of all predictions depend mostly on
the way, the Civilization will prefer : "Ahead, to HOMO SAPIENCE"
or "BACK TO PRIMATES".
\end{abstract}

\section {Introduction}
    Looking back, we can better predict our future. Few historical remarks. In Russian
    tradition, we divide the Radio history into Radio as a tool
    for "People-to People" communication and Radio as a tool for
    "People-to Nature" communication. Loomis (USA, 1876) had the first
    patent on wireless telegraph, much before Marconi and Popov,
    but Popov with his patent on "Thunderstorm marker" was the
    first in "People-to Nature" communication (Russia,
    1897).During the last
     100 years, Radio Astronomy made the really big step -- from
    the atmospherics lightning to the Big Bang...
    It was demonstrated by many radio astronomers, that, by averaging  of the small
     steps in the 5-10 Years intervals, practically all important parameters are improving
     exponentially during the whole history of Radio Astronomy (Popov point is  on the same line).

    Such big progress in Radio Astronomy tools, comparable only
    with computers science, resulted in transformation of
    "Quantitative changes" into the "Qualitative changes". Below we
    suggest few examples.

    1. From Milky Way to the End of the luminous Universe

    2. From the "FINAL PRODUCT" of the Universe activity (nearby
    objects) to the "Initial Conditions", (Inflation etc).

    3. From "Objects" to "Proto objects" (CMB anisotropy)

    4. From small addition to the Optical Astronomy to the "Only
    Window" to the Early Universe, $z>10$.

    In instrumentation:

    1. From "Receiver noise limited" devices to the "Natural
    limited" devices (Galaxy and Metagalaxy noise, CMB anisotropy
    noise etc)

    2. From Telescope dimension limits to the "Natural limits" sets
    by physical size of objects (Compton Tb limit, e.g).

    Both lists may be extended greatly.

    It is not easy to predict the main scientific targets of
    interest for future Radio Astronomy. There was nice case with
    attempt to predict Physical science targets by M.A.Markov (later- famous
     theorists in Russian Academy of Science). He made   the list of
     predictions, compiled after direct interview
     with world  best experts in Physics of the first quarter of
     20th century (including Einstein, Planck, Bohr,
     and  many others) and checked these predictions 25 years
     later- NO ONE PREDICTION WAS CONFIRMED.

     New unsolved problems with DARK Energy, Dark matter, Dark
     Ages and HE Physics will be as a long time scale targets, but
     we are going  to talk here about possible revolution in the
     FACILITIES. Few remarks.

\begin{figure}
\centerline{\includegraphics{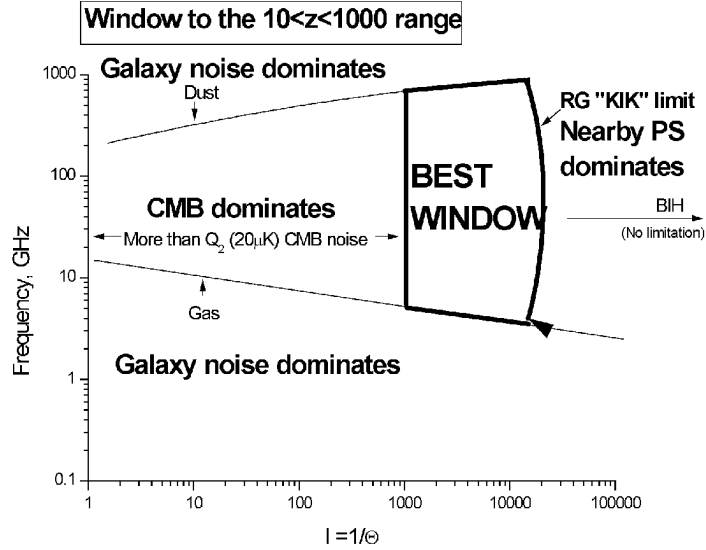}} \caption{RATAN-600 new
data on the window to the Early Universe on the "Scale --
Frequency plane"\cite{4,6}.The high l boundary depends on the
instrument parameters and on the total sky surface covered by the
extended radio sources (Ken I. Kellermann limit, marked as "KIK").
}\label{fig1}
\end{figure}

\begin{figure}
\centerline{\includegraphics{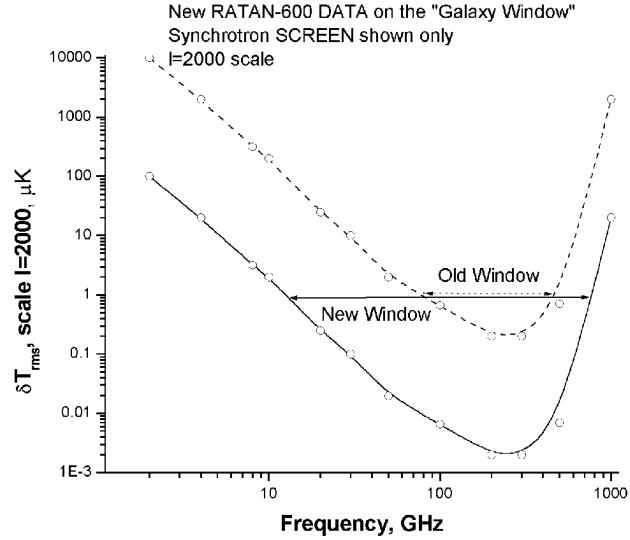}}
\caption{New limit on the Galaxy noise at l=2000
scale. Upper curve --old version, lower curve -- new, data for
synchrotron component. }\label{fig2}
\end{figure}

     1. We shall begin with the new frequency domain,
     connected with deepest observations. Human main facilities in optical
     domain(eye, telescope)
     where adjusted nicely to the maximum emission of the Sun and to the atmospheric
      window (which again was adjusted by LORD to the Sun emission). For future civilization
      we have CMB emission as a major source of energy in the Universe, and it is happened
       by some reason
      that the Universe is the most transparent to the wavelengths where CMB has maximum
      of its  emission (200GHz). New information about the width
      of this Window in the "Angular scale- frequency plane" we received recently
      during the
       3-Years long  25-frequencies monitoring of the 24 Hours in
       R.A. sky strip (\cite{4},\cite{6}).
       New window happened to be wider and deeper (see fig.\ref{fig1},\ref{fig2}), than it
       was suggested few Years ago just by strong extrapolation in
       frequency and in angular scales (\cite{7}).

 We see, that sub-microK experiments are possible in the future.

2.Phase disturbance in the wavefront, introduced by the medium
between the source under exploration and observer,  strongly
affect the information. 50 Years ago we believed, that atmospheric
phase screen set absolute limit on the resolving power of the
instruments. In Radio, we realized in 60th, that it is not the
case, and wavefront restoration is possible. Later. it was
demonstrated, that in optical domain it is possible also (Palomar
experiment with non-redundant array of the holes in the aperture
screen). Interplanetary scattering was removed also in the
properly organized observation. Recently, gravitational lensing
effects were corrected, and undisturbed map of the radio source
was found. It is very interesting, how far we can go in this way.

\begin{figure}
\centerline{\includegraphics[totalheight=7cm]{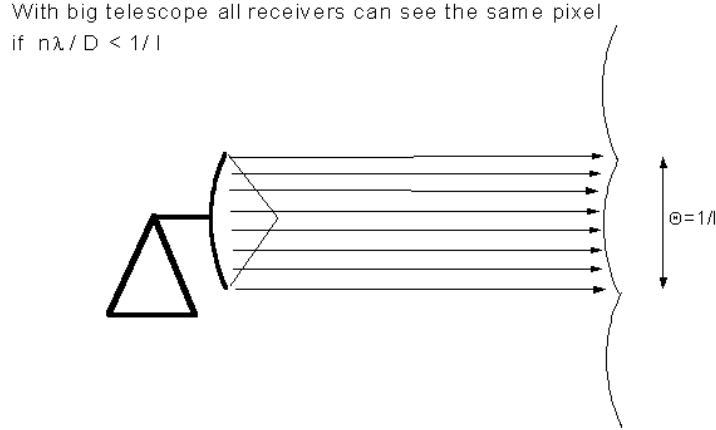}}
\caption{ If an unaberration field is smaller than the sky pixel
size of interest all focal plane receivers will see the same sky
pixel}\label{fig3}
\end{figure}

 3. Change the targets, from "Objects" to "Protoobjects",
results in
 the drastically change in requirements to the facilities. Instead of standard one (resolution,
 collecting surface etc) we need now in " temperature resolution". Up to now, there are no
 project which can work as deep as the Nature suggests. The main problem -- even with no noise
 receiver -- the integration time should be hundreds and thousands Years. One of the possible
 solution-next generation multi-elements array should be prepared not only to the registration
 $E_i*E_j$ products, but also for $(E_i)^2$, that is, to the total power mode. Looking at the same sky pixel,
 it is possible to reduce the integration time by factor N (number of elements).
    Another way, which we try to test with the world greatest reflector RATAN-600 with giant
    unaberration field -- multi-element focal plane array. Up to few thousands receivers may be
    put in the focal plane, each looking at the same pixel on the
    sky (see fig.\ref{fig3}) (\cite{5}, and "Cosmological Gene" project in www.sao.ru)

3. Up to now, the biggest in the collecting surface ground based
project was suggested in 1964 to the IAU meeting (Hamburg) by
Pulkovo radioastronomers (5 $10^6 m^2$ ring reflector array,
\cite{1}). It was too early, but now several well documented
projects with close collecting surface and with modern phase
imaging concept are under discussion. Here we want to stress, that
using simple extrapolation low, we should be ready to the $10^8
m^2 -10^9 m^2$ Radio telescopes in the middle of the 21 century.
Up to now, there are no suggestions for this Radio telescopes
class. Here we suggest two unusual (but possible) ways of
realization of $10^9 m^2$ project.

    a) From "dedicated" to "private" facilities.

     Even if someone
    is not interested in the science, but just has switched on his
    personal "People-to People" device with omnidirectional antenna (dipole, e.g.),
    he receives (see in TV mode, listening in the audio mode) all cosmic radiation above the
    horizon, including discrete objects and CMB. The level of this radiation may be from
    few percent to almost $100\%$ of the personal equipment noise.

\begin{figure}
\centerline{\includegraphics[totalheight=8cm]{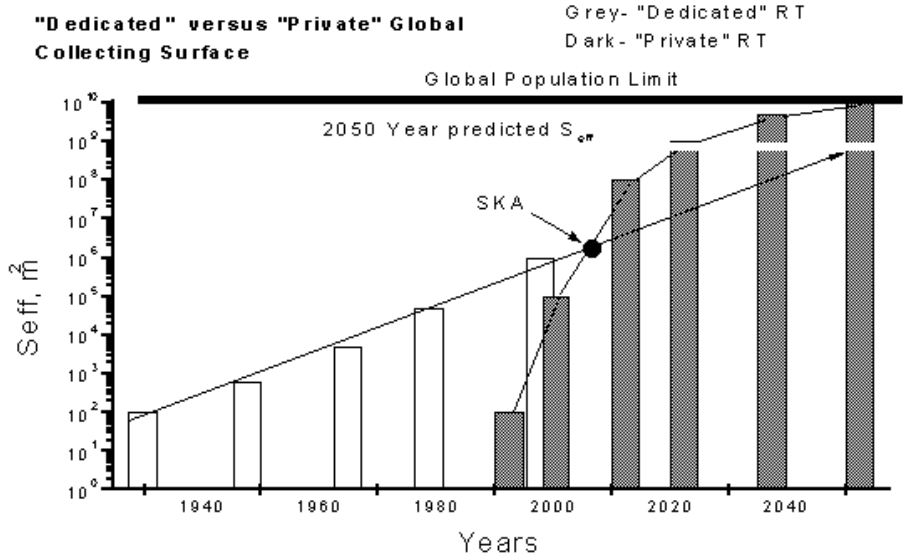}}
\caption{Global "Private" collecting surface increases much faster
than "Deducated" one for Radio Astronomy.} \label{fig4}
\end{figure}

    As is stated by the most powerful companies, in the distant
    future practically all Earth population will use personal
    People-to-People broad band communication facilities, interconnected through the global
    communication systems. With predicted 10-20 billion
    population, the collecting surface of the personal dipole type
    antenna will be not very far from $10^9 m^2$,see fig.\ref{fig4}. Phase adjustments
    may be properly introduced to the communication  signals, and
    all population may be organized as an giant phased array with big
    collecting surface, VLBI type resolution and better than VLA
    image quality (\cite{2}).

    b) Solar System resources.

\begin{figure}
\centerline{\includegraphics{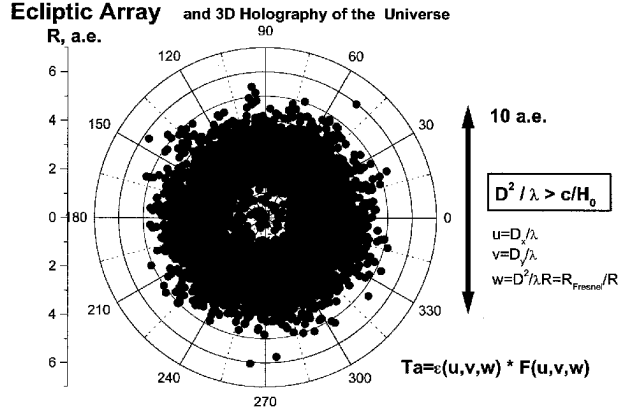}} \caption{"Ecliptic
array, using asteroids first Frehnel zones. Only 5000 objects are
shown.}\label{fig5}
\end{figure}

    As was discovered by the first generation Radar exploration of
    the Moon, the first Frehnel zone is well visible in the
    reflected signal. The same was found in other cases- from
    all Earth type planets to about 100 minor planets
    (asteroids). The size of the first zone is about $(R*\lambda)^{1/2}$ and even for minor planets
    it has about 1000 $m^2$ surface. The reflection is not ideal,
    only few percent, but it was found recently through SDSS
    survey, that number of asteroids with size
    more than 50 km increases now  to 57000 . They are scattered in the solar system inside
    the about 10 a.e. radius,see fig.\ref{fig5}. It is suggested, that the real
    number may as big as $10^8$. It means, that the total reflecting
    surface of these flying radio telescopes is much bigger, than
    SKA (\cite{2}). It is important, that this array has nano-arcsecond
    resolution, about billion $m^2$ surface, and also the whole sky
    field of view. With big dedicated radio telescope we can put first Frehnel zone
    into the near field zone, and collect the signals without distance -dependent losses.

    c) From uv- to uvw synthesis.

    With item b) type array all, deviations from the plane
    wavefront of radiation, coming from point source at $c/H_0$ (c-
    light velocity, $H_0$- Hubble constant)
    distance  may be comparable with wavelength, that is the whole
    Universe will be in the near field zone of ecliptic array. It
    means, that cosmological parallaxies  may be measured, and
    3-dimensional aperture synthesis may be realized (\cite{3}).

    d) From single pixel Radio Telescopes to the All Sky field of
    view with the same pixel sensitivity.

    It is well known, that ideal instrument should read all
    information available in the Wolf Coherent function, $W(r,\tau)$,
    where r is the space distance between two points on the
    collecting surface and $\tau$ is the time lag between these
    signals. Space structure and frequency spectrum may be collected
     from this function. The problem is, that reflectors can solve this
     problem only at the optical axes,  that is ONE PIXEL
     solution may be found. Aperture  synthesis array are better,
     but just 1/GAIN field of view may be realized
     (GAIN=Physical array element surface$/(4\pi \lambda^2)$. We hope,
     that by some way the field of view problem will be solved,
     and all sky may be mapped with greater sensitivity  than achieved now with VLA in the
     one-arcmin. field.

\section {Conclusion}

    We predict the "Second Birth" of the Radio Astronomy, and not
    only due to expected e-fold increase of the facilities
    potential, but also due to  drastically change in the role  it will
     play in the future; from secondary (to optical domain) to the
     "prima ballerina" role at $z>10$

\end {document}